# LINEAR POLARIZATION OF THE PHOTOLUMINESCENCE OF QUANTUM WELLS

A.V. Koudinov[†], N.S. Averkiev, Yu.G. Kusrayev, B.R. Namozov and B.P. Zakharchenya

*A.F. Ioffe Physico-Technical Institute, 194021 St. Petersburg, Russia*

D. Wolverson and J. J. Davies

*Department of Physics, University of Bath, Bath BA7 2AY, UK*

T. Wojtowicz, G. Karczewski and J. Kossut

*Institute of Physics, Polish Academy of Sciences, 02-668 Warsaw, Poland*

The degree and orientation of the magnetic-field induced linear polarization of the photoluminescence from a wide range of heterostructures containing (Cd,Mn)Te quantum wells between (Cd,Mn,Mg)Te barriers has been studied as a function of detection photon energy, applied magnetic field strength and orientation in the quantum well plane. A theoretical description of this effect in terms of an in-plane deformation acting on the valence band states is presented and is verified by comparison with the experimental data. We attempted to identify clues to the microscopic origin of the valence band spin anisotropy and to the mechanisms which actually determine the linear polarization of the PL in the quantum wells subject to the in-plane magnetic field. The conclusions of the present paper apply in full measure to non-magnetic QWs as well as ensembles of disk-like QDs with shape and/or strain anisotropy.

## I. Introduction

The linearly polarized luminescence of semiconductor nanostructures can yield information on the details of their symmetry, on the mechanisms of the inter-particle exchange interaction and on the interaction of the particles with external fields. For quantum wells (QWs) and disk-like quantum dots, it also contains valuable information on the confinement potential in the lateral direction, which cannot be obtained from conventional (polarization-independent) spectroscopy. In Ref. [1], an unusual behaviour of the degree of linear polarization of the emission of CdTe/(Cd,Mn)Te QWs was discovered with a magnetic field parallel to the QW layer. The most remarkable finding of Ref. [1] was the extremely anisotropic g-factor of the valence band states and this was explained as a consequence of a uniaxial in-plane distortion lowering the QW symmetry; an important contribution to the theory has been presented recently Ref. [2]. The concept of an extremely anisotropic (but *pseudo-isotropic*) hole g-factor has been verified by spin-flip Raman scattering (SFRS) experiments [3] and photoluminescence (PL) studies of charged single quantum dots [4], where the energy separation of the valence band spin sublevels was spectrally resolved.

The present paper is devoted to the detailed investigation of the contributions to the linear polarization of the PL of QWs and aims at the generalization of results obtained on many samples, and at the understanding of the factors controlling the PL polarization in QWs of (for example) different thicknesses, barrier heights, or concentrations of magnetic ions. Attention has been paid, both in theory and in experiment, to the analysis of the contributions to the polarization dependence having different symmetries (the "angular harmonics" of polarization) and the complementary techinques of PL and SFRS have been applied. The theory developed is compared to the experimental angular, magnetic-field and spectral dependences of the polarization.

---

[†] koudinov@orient.ioffe.ru

The paper is organized as follows. In Section II we describe the samples under study and give a brief description of the experimental techniques. Section III contains the review of our experimental results. In Section IV we develop the theoretical model. Section V is devoted to a comparison of the theoretical and experimental results. Finally, Section VI lists the main results and conclusions of the paper.

## II. Samples and experimental details

We have studied several samples and present here results on six QW heterostructures that provide typical examples of the polarization behavior of interest; their details are summarized in table 1.

| Sample number | $Cd_{1-x}Mn_xTe$ quantum well: x. | $Cd_{1-x-y}Mn_xMg_yTe$ barrier: x, y. | Quantum well widths (Å) | Barrier width (Å) | Buffer type | Substrate |
|---|---|---|---|---|---|---|
| 1 | 0.00 | 0.10, 0.00 | 40, 60, 80 | 500 | A | (001) GaAs |
| 2 | 0.00 | 0.30, 0.00 | 20, 40, 60, 80 | 500 | A | (001) GaAs |
| 3 | 0.00 | 0.50, 0.00 | 20, 40, 60, 80 | 500 | A | (001) GaAs |
| 4 | 0.00 | 0.30, 0.00 | 20, 40, 60, 80 | 500 | B | (001) GaAs |
| 5 | 0.07 | 0.07, 0.29 | 30 | - | C | (001) GaAs |
| 6 | 0.07 | 0.07, 0.20 | 9, 16, 45, 80, 300 | 500 | - | (001) $Cd_{0.964}Zn_{0.036}Te$ |

Table 1. The details of the samples used in this work. Buffer type A consisted of 0.2 μm ZnTe, 0.8 μm CdTe and 2 μm $Cd_{1-x}Mn_xTe$. Type B contained an "aperiodic superlattice" (ASL) of 10 alternating layers of ZnTe, CdTe and (Cd,Zn)Te of thickness 3-20 nm, followed by a 3 μm layer of CdTe and a 0.7 μm layer of (Cd,Mn)Te; type C consisted of a 6 μm CdTe layer.

In the PL experiments, the magnetic field was in the plane of the QWs ($H \perp [001]$) and the luminescence collected was propagating normal to the plane (the Voigt configuration). The PL was excited by either helium-neon or argon ion lasers, with a pump density of less than 1 W/cm$^2$. The linear polarization data were taken using a photo-elastic quartz polarization modulator and a two-channel photon counting technique. As the excitation energy was well above the recombination energy, *no influence* of the excitation polarization on the PL polarization was detected in our experiments.

To obtain complete information about the linear polarization of the PL, we measured the two polarization parameters defined as

$$\Re_0 = (I_\perp - I_\parallel)/(I_\perp + I_\parallel), \quad \Re_{45} = (I_{+45} - I_{-45})/(I_{+45} + I_{-45}),$$

where $I_\perp$ and $I_\parallel$ stand for the intensities of PL polarized perpendicular and parallel to the magnetic field, respectively, while $I_{+45}$ and $I_{-45}$ represent those polarized in two orthogonal directions rotated (in the sample plane) by 45° with respect to the field. These two parameters are unambiguously related to the true (total) polarization degree and the direction of the plane of polarization [1]. The *angular scan*s of polarization (i.e., dependences of the polarization degrees $\Re_0$ or $\Re_{45}$ on the angle $\varphi$ which the magnetic field makes with the [110] axis of the crystal) were obtained by rotating a sample immersed in superfluid helium (T ~ 2K) around its growth axis.

In the SFRS experiments, a Ti-sapphire laser pumped by the green/blue output from an argon ion laser was used to provide resonant excitation and the scattered light was analyzed in a spectrometer with a double subtractive filter stage followed by a final dispersing stage of focal length 1 meter. The light was detected either with a charge coupled detector array or with a cooled GaAs photomultiplier photon counting system. The specimens were mounted in direct contact with superfluid helium at 1.6 K in a superconducting magnet that provided fields up to 6 Tesla and the Raman spectra were taken in the back-scattering mode. The specimen orientation relative to the field direction (horizontal) could be changed by rotation about the vertical axis by an angle $\theta$, so that $\theta=0°$ and $\theta=90°$ correspond to the Faraday and Voigt configurations respectively.

## III. Experimental results

III.1 Polarized photoluminescence

Fig. 1 shows the typical PL spectra of the CdTe/(Cd,Mn)Te QWs. The characteristic feature of such spectra is the presence of the two lines separated from each other by about 4 meV (Fig. 1(a,b). Depending on the properties of a given sample, and on the QW thickness, these lines may be well-resolved or may overlap, as seen in Fig. 1(c). The ratio of the intensities of the high- and low-energy lines is also influenced by the temperature, the magnetic field, the energy of the exciting photons and the pump density [5].

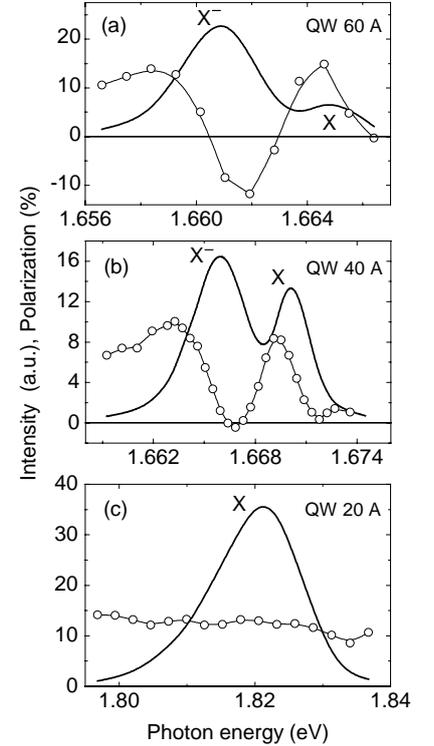

Fig.1 Photoluminescence (solid lines) and linear polarization degree $\Re_0$ (circles connected by a spline) spectra taken from different QWs in the magnetic field $\boldsymbol{H}\perp z$: (a) sample 2, QW 60Å, $H$=2.5 T; (b) sample 1, QW 40Å, $H$=0.4 T; (c) sample 4, QW 20Å, $H$=1.5 T. Excitation by He-Ne laser (1.96 eV), T=2K.

The interpretation of the lines is as follows. The high-energy line (X) is due to recombination of excitons formed by electrons and holes from the lowest-energy QW levels for the respective energy bands (*1e-1hh* excitons in common notation). The excitons responsible for the X line are the quasi-2D excitons localized at the interface roughnesses. The low-energy line is also formed by the *1e-1hh* excitonic states but arises either from negatively charged excitons, *trions* ($X^-$), or from excitons localized on neutral donors ($D^0X$). It is not easy to distinguish experimentally between the two possibilities [6], as both the trion and the donor complex include one hole and a pair of electrons [7]. In what follows we will refer to the low-energy line as the trion one, whilst bearing in mind that every conclusion presented here for trions would apply equally well to donor-bound excitons.

Fig. 1 also shows the spectral dependence of the PL polarization parameter $\Re_0$ at the magnetic field applied. The energy dependence of the polarization degree has a remarkable shape; the polarization degree shows a quadratic increase with magnetic field (in the low-field domain) and, as was discovered earlier, depends strongly on the orientation of the magnetic field with respect to the crystallographic axes.

An analysis of the influence of symmetry on the properties of the PL polarization for various QWs is of great interest. As shown in Ref. [1], for an ideal [001]-oriented QW possessing $D_{2d}$ symmetry, the angular scans $\Re_0(\varphi)$ can contain only zeroth ($\Re_0$ is a contant) and fourth ($\cos 4\varphi$) angular harmonics. However, experiments showed that, in fact, the second angular harmonic ($\cos 2\varphi$) also contributes, its amplitude often being larger than those of zeroth and fourth harmonics. It was concluded that the true symmetry of all the QWs studied in Ref. [1] was orthorhombic ($C_{2v}$), so that the axes [110] and [1$\bar{1}$0] lying in the QW plane were not equivalent; $C_{2v}$ symmetry allows zeroth, second and fourth harmonics in the quadratic approximation in field. All these harmonics had been discovered in the angular scans. The ratios of their amplitudes were different for different QWs.

The most striking results of the angular scan are shown in Fig. 2(a), which shows that the second angular harmonic alone is sufficient to fit the data for $\Re_0$. This implies that, for any orientation of the magnetic field in the plane of the QW, the predominant polarization of the PL is unchanging in direction and is directed, depending on the detection energy, along either the [110] or [1$\bar{1}$0] axis (see Fig. 1(a)). The total polarization degree depends on the strength but not on the direction of the magnetic field. Measurement shows that, even at *zero* magnetic field, a weak linear polarization (of a few percent) directed along [110] is observed (referred to in what follows as the "built-in polarization"). The angular scan of the built-in polarization displays also the second harmonic (Fig. 2(b)); this fact, however, is only the trivial consequence of the rotation of the QW with respect to the laboratory reference frame in which the intensities $I_\perp$ and $I_\parallel$ are defined. Thus, the magnetic field-induced polarization of the PL demonstrates exactly the same behaviour as the built-in polarization (cf. Fig. 2(a) and 2(b)), that is, the former is in no way related to the orientation of the magnetic field though being induced by the field.

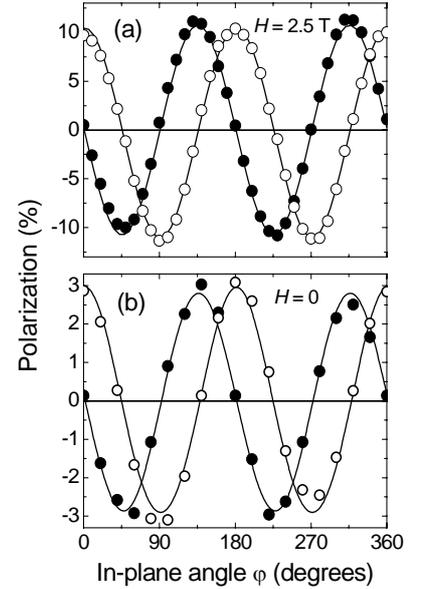

Fig.2 In-plane angular scans of polarization parameters $\Re_0$ (open circles) and $\Re_{45}$ (closed circles) taken in the following conditions: (a) sample 2, QW 60 Å, $H$=2.5 T (***H***⊥*z*); (b) sample 6, QW 45 Å, $H$=0. Angle $\varphi$=0° corresponds to ***H***∥[110]. Lines show the pure second harmonic fits: $\Re_0 \propto \cos 2\varphi$, and $\Re_{45} \propto \sin 2\varphi$.

In Ref.[1] an explanation of the angular scans similar to that shown in Fig. 2(a) was suggested on the basis of an extremely anisotropic lateral g-factor of holes, induced by a low-symmetry distortion of the QW. The microscopic origin of the distortion remained obscure (possibilities include a uniaxial deformation in the plane of the QW, anisotropy of the single heterointerface, or an anisotropy of the "islands" localizing the quasi-2D excitons). The following results help shed some light on the origin of the reduced symmetry in these samples.

Fig. 3(a) depicts the angular dependences of $\Re_0$ and $\Re_{45}$ for the narrow QW in sample 2. Fitting shows that here, in contrast to Fig. 2(a), the contributions of all three symmetry-allowed harmonics are present in the angular scan of $\Re_0$. The amplitudes of the zeroth and second harmonics have comparable values, while that of fourth harmonic is smaller (the fourth harmonic manifests itself in the sharpening of the maxima and the flattening of the minima.). Fig. 3(b) shows the angular scans for the $Cd_{0.7}Mn_{0.3}Te$ barrier layer in the same sample. The anisotropy of $\Re_0$ in the barrier is similar to that in the QWs, except that the fourth harmonic is more pronounced. The anisotropy of the magnetic field-induced linear polarization of the PL in bulk crystals has been reported earlier [8]. When the light is emitted along [001], the symmetry of the bulk crystal allows only zeroth and fourth harmonics in the angular scan. Here, we are principally interested in the fact that the angular scans of the barrier PL clearly contain the second angular harmonic, i.e., not only the QWs but also the barriers are distorted in their plane.

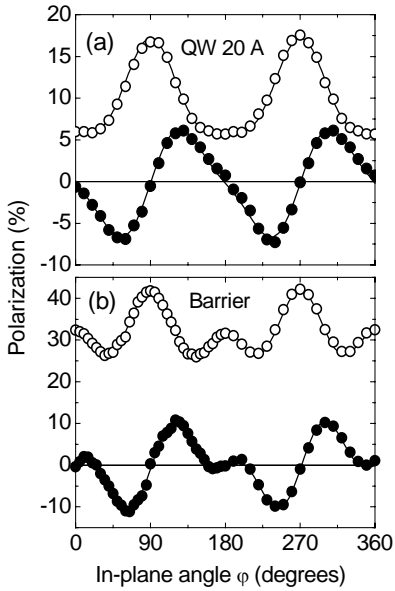

Fig.3 In-plane angular scans of polarization parameters $\Re_0$ (open circles) and $\Re_{45}$ (closed circles) taken for: (a) sample 2, QW 20 Å, $H$=2.5 T; (b) sample 2, luminescence from the barrier layer, $H$=1 T. Points are fitted by a sum of zero-th, second and fourth harmonics (in case of $\Re_0$) and by that of second and fourth harmonics (in case of $\Re_{45}$).

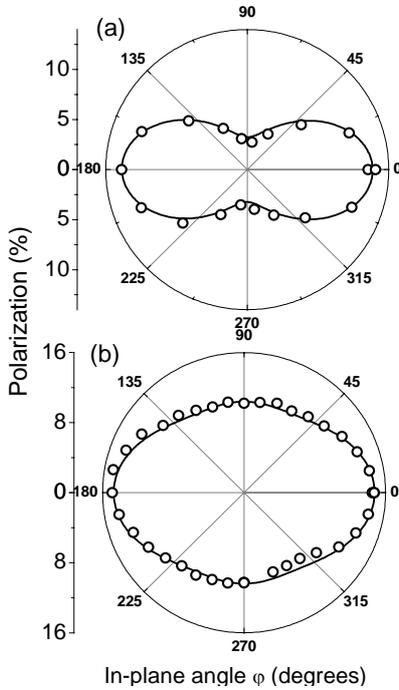

Fig.5 In-plane angular scan of polarization parameter $\Re_0$ (shown in polar coordinates) taken for: (a) sample 6, QW 45 Å, $H$=1.25 T; (b) sample 5, QW 30 Å, $H$=2.6 T.

Fig. 4(a,b) presents the results of the measurements on sample 4 under analogous conditions to those of Fig. 3. This sample was grown with the same design as the sample 2 except for an additional ASL buffer that was deposited in order to stop misfit dislocations spreading into the heterostructure during growth. One can see that in Fig. 4 the contribution from the second harmonics is much smaller than in Fig. 3. Thus we are led to the conclusion that, in samples with a "conventional" buffer, the $C_{2v}$ distortion comes from the lattice misfit of the substrate and the heterostructure. One should note that, even for sample 4, the angular scans of the wider QWs demonstrate more clearly the presence of the second harmonic, though it is less pronounced than in QWs of the corresponding thickness in sample 2. Therefore, the ASL buffer is not completely effective in permitting relaxation of the misfit.

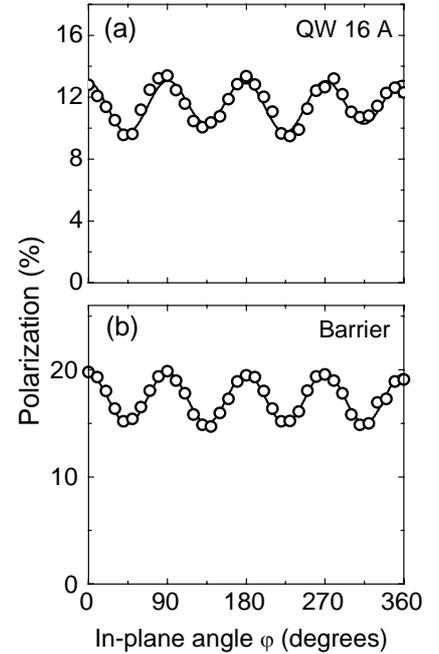

Fig.4 In-plane angular scans of polarization parameter $\Re_0$ (open circles) taken for: (a) sample 6, QW 16 Å, $H$=1.5 T; (b) sample 6, luminescence from the barrier layer, $H$=0.3 T.

The angular scans of the (Cd,Mn)Te/(Cd,Mg,Mn)Te QWs, in which the Mn ions are inserted both in the barrier and in the QW layers, are qualitatively similar to those of CdTe/(Cd,Mn)Te QWs. Fig. 5 presents the results obtained on samples 5 and 6. The angular scan of sample 6 (Fig. 5(a)) is dominated by the second harmonic. However, one should note that in the QWs of sample 6, in contrast to the QWs of all other samples, the amplitude of the built-in polarization is noticeable, so the major part of the second harmonic seen in the figure is the zero-field effect. Sample 6 has been grown under a different set of conditions to samples 1-5 (in particular, on a different substrate). Meanwhile, the polarization scan of sample 5, which was grown under the same conditions as the majority of the samples, is dominated by the zeroth harmonic, though the second and the fourth are also present.

We now discuss the increase of the PL polarization as the magnetic field is increased. Fig. 6(a) depicts the dependences of $\Re_0$ on the value of the magnetic field for the two orientations of the field – when it is parallel to the [110] axis (along the in-plane distortion) and to the [100] (at 45° to the distortion). The angular scan of the present QW is dominated by the second harmonic, but the zeroth harmonic also admixes. Therefore, the increase of the polarization when the field is along [110] mainly corresponds to the increase of the amplitude of the second harmonic. One can see that, above ~ 0.5 T, the initially quadratic increase of polarization with field becomes weaker. For a field along [100], the second harmonic obviously cannot contribute to $\Re_0$ (since cos 2φ equals zero), so the latter dependence reflects the increase of the zeroth harmonic. Here, the polarization again increases quadratically, but with a smaller coefficient. However, the increase of polarization remains quadratic over the whole field range of 2.5 T.

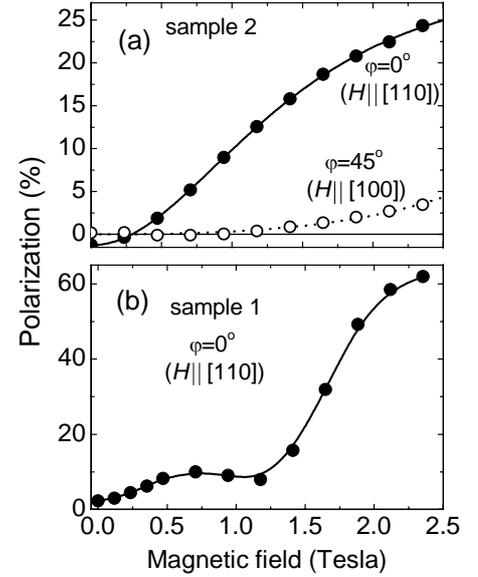

Fig.6 Magnetic field dependences of the polarization degree $\Re_0$ taken for: (a) sample 2, QW 40 Å, two different orientations of the crystal with respect to the field; (b) sample 1, QW 40 Å. Solid lines are guides for the eye. Dotted line in panel (a) shows the quadratic

Another interesting case of the field dependence of $\Re_0$ is shown in Fig. 6(b). This curve was recorded for sample 1, a structure with rather low barriers and a high effective concentration of Mn ions [9] and thus with an especially strong exchange enhancement of the external magnetic field. Here, one can see a complex and non-monotonic behaviour of the polarization: a quadratic growth is replaced by a kind of saturation, followed by a new region of polarization increase.

Fig.7 demonstrates the field dependences of the amplitudes of the three angular harmonics in one QW; a QW was chosen for which the angular scan of $\Re_0$ contained a dominant zeroth harmonic with smaller but significant second and fourth harmonics, so that a comparison of the behaviour of all the harmonics was possible. We have measured the angular scans of $\Re_0$ for a selection of magnetic field values. Each angular scan was then fitted by a sum of harmonics whose amplitudes were the fitting parameters. We also have checked that each dependence was an even function of field. Although for second and fourth harmonics, the amplitudes do not exceed 3%, the data points lie well on a smooth trend. In fact, the experimental error for each point in Fig.7 is smaller than that of a single polarization measurement, since each value of the amplitude was obtained from many single measurements making the angular scan. The results show that the zeroth and the second harmonics are approximately quadratic in the whole field range up to 2.5 T whereas, for the fourth harmonics, the points lie on a straight line in the range 0.25 to 1.75 T.

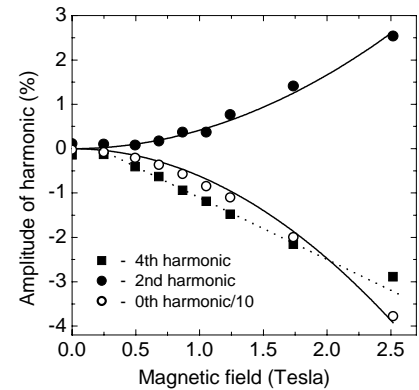

Fig.7 Magnetic field dependences of the amplitudes of zeroth, second and fourth angular harmonics of $\Re_0$, taken for sample 6, QW 16 Å. Solid lines show the quadratic fits of the two former dependences. Since for the fourth harmonic the quadratic fit does not work in the chosen range of the magnetic fields, the straight line is given instead (dotted).

## III.2 Spin-flip Raman scattering

While the PL polarization measurements are highly sensitive to weak distortions of the QW symmetry, they are not well suited for quantifying the electron and hole spin splittings. The measured value of the PL polarization degree is affected by both the electron and hole spin polarizations. Moreover, in diluted magnetic semiconductors, the spin polarizations are often controlled not so much by the simple ratios of the corresponding spin splitting and the temperature, but by more specific factors as the magnetic polaron effect, magnetic fluctuations, etc. (see [10] and subsection V.5)

Spin-flip Raman scattering measurements allow one to obtain more direct information about the splitting of the electron and hole spin sublevels in a magnetic field. The spin-flip Raman spectra of QWs demonstrating a pseudo-isotropic hole *g*-factor were studied recently [3]. That work focussed on spectra obtained in the Voigt configuration; here, we present results for a tilted magnetic field making a range of angles $\theta$ to the growth axis and we confirm the validity of some of the approximations that will be made in the theoretical treatment of Sect. IV.

Fig. 8 demonstrates a series of spin-flip Raman spectra taken at $\theta = 60°$ with values of the magnetic field from 1 to 6 Tesla. The optical excitation was in resonance with the X excitons in the QW (see Fig.1a). The sharp line close to the laser (marked MnSF) corresponds to the spin flip of electrons in the 3d shells of manganese ions and the line marked ESF arises from the spin flip of band or donor-bound electrons [3]. The two lines marked HSF and XSF are absent in the pure Voigt configuration; as the magnetic field increases, the Raman shifts of the HSF and the XSF lines tend to saturation, similar to that of ESF and in contrast to that of Mn, as shown in Fig. 9(a). This implies that the HSF and XSF lines are due to the spin flips of the charge carrier band states which experience the action of the effective "exchange field" of the magnetic ions. Similar lines have been observed also in SFRS studies of related non-magnetic QW systems (for instance, by Sirenko et al. [11]), where a signal similar to XSF was ascribed to an exciton spin flip process; this identification is confirmed below.

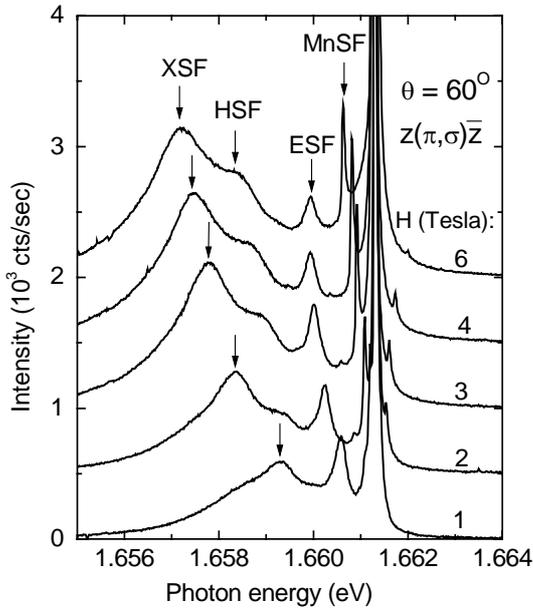

Fig.8 Resonant spin-flip Raman scattering in sample 2, QW 60 Å subject to tilted magnetic field ($\theta=60°$ to the growth axis) at different strength of the field. Spectral position of the laser is seen in the figure, T=1.6 K, spectra are consequently upshifted for convenience. Spectral features corresponding to spin flips of manganese, electron, hole and exciton are marked as described in the text.

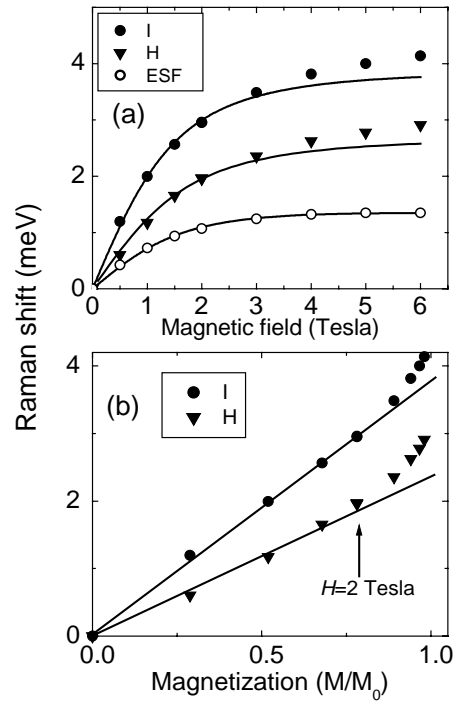

Fig.9 Raman shifts of the ESF, HSF and XSF features in sample 2, QW 60 Å (at *H* along $\theta=60°$) plotted versus the value of the field (panel *a*) and the magnetization of the manganese spin system (panel *b*). Lines in panel *a* show the Brillouin fits.

The field-dependence of the ESF Raman shift is reasonably well described by a Brillouin function [9] plus the intrinsic conduction band Zeeman splitting with the following parameters: effective temperature $T + T_0 = 3.8$K (higher than the bath temperature of 1.5K because of antiferromagnetic coupling effects at the high Mn concentration of the barrier [12]); effective paramagnetic Mn concentration experienced by the quantum well electrons $x = 0.0034$; s-d exchange constant $N_0\alpha = 0.22$ eV, and conduction band g-factor for the quantum well of -1.3; this latter is altered from the intrinsic CdTe conduction band g-factor value of -1.67 [13] because of quantum confinement effects [11,14]. Several alternative sets of fitting parameters can be found but we have chosen the set most consistent with the previous work cited here. Interestingly, the Brillouin fit works less well for HSF and XSF because of their slower saturation, as demonstrated by Fig. 9(b), where the Raman shifts of HSF and XSF are plotted as a function of magnetization taken from the Raman shift of ESF. As the magnetization increases, the spin splittings first increase linearly and then acquire a superlinear contribution. We note that the slower saturation of HSF and XSF could in principle arise from an intrinsic valence band Zeeman splitting $g_h\mu_B H$ but a fit including such a term to this data requires $g_h \sim 2.1$ to 2.2, which is implausibly large [11,14].

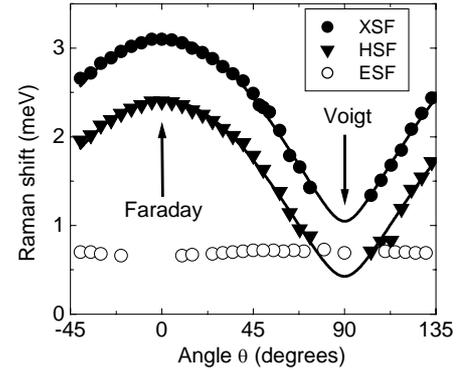

At a fixed magnetic field, the Raman shifts of the Mn and ESF lines do not depend on $\theta$, since the g-factors of the 3d electrons of manganese and of the conduction band electrons are both to a good approximation isotropic [11]. On the other hand, the Raman shifts of the HSF and XSF depend strongly on $\theta$, as shown in Fig.10. This confirms that the Raman shifts of both processes include contributions from the heavy hole g-factor, which is expected to possess a longitudinal component $g_\parallel$ (along the growth axis) much larger than the in-plane components [15]. The sum of the Raman shifts of the ESF and HSF equals the Raman shift of XSF and so we attribute the HSF signal to the spin flip of the hole and XSF to the correlated spin flip of the electron and hole of an exciton [16,17]. The $\theta$-dependences of the HSF and XSF Raman shifts are well described by the formula

Fig.10 Out-of-plane angular dependences ($\theta$-dependences) of the Raman shifts of ESF, HSF and XSF features corresponding to spin flips of the electron, the hole and of the exciton, respectively. $H=1$ T, $T=1.6$ K. Lines show the results of calculation as described in Section *III.2*.

$$\Delta_h = \mu_B H \sqrt{g_\parallel^2 \cos^2\theta + g_\perp^2 \sin^2\theta},$$

(when, for XSF, the angle-independent electron spin flip value is also added). The observed ratio of $g_\parallel$ to $g_\perp$ (5.7±0.7) is in good agreement with the values obtained for the same QW in Ref. [3] by the analysis of the Raman excitation profiles (5.7±1.5) and from the angular dependences of the excitonic luminescence (5.6±0.7).

These results allow us to define, for the purposes of the following theoretical discussion, the extent of the low-field region in which the electron and hole splittings are proportional and are dominated by the term due to the exchange field of the Mn ions. This region depends on the sample and, for sample 2, extends up to about 2 Tesla, as indicated on Fig. 9(b).

**IV. Theory**

In rectangular QWs based on cubic semiconductors, the ground state of holes is the heavy hole state characterized by a projection of the angular momentum onto the growth axis [001] equal to ±3/2 (we define $z \parallel$ [001] while $x$ and $y$ axes lie in the plane of the QW). Therefore, for an in-plane magnetic field, there is no splitting of the ground state in a linear approximation in the field ($g_\perp \sim 0$ and so $\Delta_h = 0$ for $\theta = 90°$). This calls for specific mechanisms responsible for the linear polarization of radiation. The present section presents a detailed investigation of those mechanisms.

The selection rules for radiative transitions between the electrons in the $\Gamma_6$ state and the holes in the $\Gamma_8$ state have the form [15]

$$\left\langle \alpha \left| \frac{3}{2} \right\rangle \right. = -A \frac{e_x + i e_y}{\sqrt{2}}; \quad \left\langle \alpha \left| \frac{1}{2} \right\rangle \right. = A \sqrt{\frac{2}{3}} e_z; \quad \left\langle \alpha \left| -\frac{1}{2} \right\rangle \right. = A \frac{e_x - i e_y}{\sqrt{6}}; \quad \left\langle \alpha \left| -\frac{3}{2} \right\rangle \right. = 0;$$
$$\left\langle \beta \left| \frac{3}{2} \right\rangle \right. = 0; \quad \left\langle \beta \left| \frac{1}{2} \right\rangle \right. = -\left\langle \alpha \left| -\frac{1}{2} \right\rangle \right.^*; \quad \left\langle \beta \left| -\frac{1}{2} \right\rangle \right. = \left\langle \alpha \left| \frac{1}{2} \right\rangle \right.; \quad \left\langle \beta \left| -\frac{3}{2} \right\rangle \right. = -\left\langle \alpha \left| \frac{3}{2} \right\rangle \right.^*$$
(1)

where $A$ is a constant, $e_j$ is the electric dipole transition amplitude for a light polarization vector in direction $j=(x, y, z)$, and $\alpha$ and $\beta$ denote the electron states with $z$-projection of the spin equal to +1/2 and −1/2 respectively. Eqs.(1) show that, to obtain linearly polarized ground-state emission propagating in the $z$ direction, it is necessary to admix the hole states |-1/2⟩ and |+3/2⟩ or |+1/2⟩ and |-3/2⟩; the recombination of an electron with such a hole state generates a coherent superposition of two photons having opposite circular polarizations and, therefore, a linear component to the emission.

Mixing of states |+3/2⟩ with |-1/2⟩ or |-3/2⟩ with |+1/2⟩ can be accomplished either by an in-plane magnetic field (in second order perturbation theory) or by a uniaxial in-plane deformation of the QW (in first order perturbation theory). In the former case, the linear polarization is quadratic in the magnetic field and does not depend on the electron spin orientation. In the latter case, the polarization is linear in the deformation and does not require application of the magnetic field. Finally, if both a magnetic field and a deformation are imposed, there arises a splitting of the heavy hole levels which is linear in both the field and the deformation. This results in a linear polarization of the emission which depends on the spin orientation of electrons and holes. Let us consider in more detail the mechanisms responsible for the linear polarization.

We assume that recombination takes place between the electron and the hole with momentum **k**=0, while the magnetic field **H** lies in the plane of the QW. The Hamiltonian describing the splitting of the electron states may be presented as

$$H_e = g_e (\mathbf{sH}),$$
(2)

where $g_e$ stands for the g-factor of the electron[§] and **s** is the spin operator. The splitting of the electron levels, $\Delta E_e = g_e H$, does not depend on the orientation of the magnetic field. The Hamiltonian describing the splitting of the hole states in the simplest approximation, when only the ground state and the first excited state of the hole are accounted for, has the form

$$H_h = \begin{pmatrix} 0 & \frac{\sqrt{3}}{2}g_1 H_- & \frac{\sqrt{3}}{2}\eta & 0 \\ \frac{\sqrt{3}}{2}g_1 H_+ & -\Delta & g_1 H_- & \frac{\sqrt{3}}{2}\eta \\ \frac{\sqrt{3}}{2}\eta & g_1 H_+ & -\Delta & \frac{\sqrt{3}}{2}g_1 H_- \\ 0 & \frac{\sqrt{3}}{2}\eta & \frac{\sqrt{3}}{2}g_1 H_+ & 0 \end{pmatrix}. \quad (3)$$

Here $H_\pm = H_x \pm i H_y$, $\eta$ is the value of the in-plane deformation multiplied by the respective constant of the deformation potential, $\Delta$ is the energy separation between the heavy and the light holes, $g_1$ is the hole g-factor for the bulk material [15], and the principal axis of the in-plane deformation is taken as the $x$ axis.

An important note is that, as can be seen from Eq.(3), the deformation and the magnetic field mix heavy hole states only with the light hole ones, whereas the light hole states are mixed both with light and heavy hole states. That is why the Hamiltonian (3) can be considered as an effective Hamiltonian for the ground state holes in the QW, in which all the excited states with projections of the angular momentum $|\pm 1/2\rangle$ are taken into account by the phenomenological parameters $g_1$ and $\eta$. Furthermore, the value of the matrix elements $H_{12}$ and $H_{23}$ may differ not only by the numeric coefficient $\sqrt{3}/2$, but also by some function of effective masses, barrier height, etc. For this reason we will write $H_{12} = \sqrt{3} g_1 H_- / 2$, $H_{23} = g_1^* H_-$, where

$$\begin{aligned} g_1 &= g \int_R \psi_{lh}(z) \psi_{hh}(z) dz \\ g_1^* &= g \int_R \psi_{lh}^2(z) dz \\ g_1^{long} &= g \int_R \psi_{hh}^2(z) dz \\ g &= \frac{7}{6} x \frac{\beta N_0}{kT} g_0 \mu_B \end{aligned} \quad (4)$$

Here $g_0$ is the g-factor of $3d^5$-electron of Mn ion, $x$ stands for the "effective" concentration of Mn ions in the barriers, $\mu_B$ is the Bohr magneton, $\beta N_0$ is the exchange interaction constant of holes with $3d^5$ electrons [9], and $\psi_{lh}$ and $\psi_{hh}$ are the envelope functions of light and heavy holes respectively. The integration in (4) is performed over the regions $R$ containing the Mn ions, and $\psi_{lh}$ and $\psi_{hh}$ are normalized. The expression for $g_1^{long}$, which represents the effective g-factor of heavy holes in the Faraday geometry, will be used in the next Section.

---

[§] In what follows we imply that the Bohr magneton is included in the definition of the *g*-factors.

Calculation shows that in linear approximation in both $H$ and $\eta$, a splitting of the heavy hole levels $\Delta E_h = 3g_1 H\eta/\Delta$ appears, where $\Delta$ is the subband separation and is assumed to exceed all splittings caused by external fields and to satisfy $\Delta >> kT$. In order to explain our experimental results on the PL polarization, we need to determine the wave functions and the energy levels of the hole ground state to order $H^2$. A convenient way to do this is, firstly, to find out the eigenvalues and the eigenfunctions of Eq.(3) at $H = 0$ and, secondly, to carry out the calculations to order $H^2$ inclusive. As the experimental situation satisfies the condition $\eta/\Delta << 1$, we shall retain the terms to up to order $\eta^2/\Delta^2$ in the expressions for the light intensities.

Assume that in equilibrium the populations of the hole states are proportional to $\exp(E_h/kT)$, while those of electron states – to $\exp(-E_e/kT)$. If the inequalities $\eta g_1 H/kT << 1$ and $g_e H/kT << 1$ are fulfilled (the case of weak thermal orientations of holes and electrons), then the expression for the degree of linear polarization of the PL may be presented in the form

$$\Re_0 = \wp \cos 2\phi + A_0 H^2 + A_2 H^2 \cos 2\phi + A_4 H^2 \cos 4\phi , \qquad (5)$$

where $\phi$ is the angle between the magnetic field and the principal axis of the deformation, and $A_i$ and $\wp$ are given by Eq. 6

$$A_0 = \frac{g_1 g_1^*}{\Delta^2}\left(1 - \frac{19}{2}a_1^2\right) + \frac{2a_1^2 g_1 g_1^*}{\Delta kT}$$

$$A_2 = \frac{\sqrt{3}}{3}\frac{a_1}{\Delta^2}\left(g_1^{*2} - 5g_1^2\right) - \frac{\sqrt{3}g_e g_1 a_1}{(kT)^2} + \frac{2}{\sqrt{3}}\frac{g_e g_1 a_1}{\Delta kT} + \sqrt{3}\frac{a_1 g_1 g_1^*}{\Delta kT}$$

$$A_4 = -\frac{g_e g_1^*}{\Delta^2}\frac{19}{12}a_1^2 - \frac{g_e g_1^* a_1^2}{2(kT)^2} + \frac{4}{3}\frac{g_e g_1^* a_1^2}{\Delta kT} + \frac{g_e g_1^* a_1^2}{2\Delta kT} \qquad (6)$$

$$a_1 = \frac{\sqrt{3}}{2}\frac{\eta}{\Delta}, \quad \wp = \frac{\eta}{\Delta}$$

In making comparison between the $A_i$, it will be significant that $A_0$ and $A_4$ depend on $a_1^2$ whereas $A_2$ is linear in $a_1$ and, therefore, in the deformation. The first term in $A_0$ determines the PL polarization *independent* of the spin polarization of holes (analogous to van Vleck paramagnetism).

The opposite extreme is realized under the conditions that the splittings of both electron and hole states exceed $kT$: $\eta g_1 H/kT >> 1$ and $g_e H/kT >> 1$; then, only one optical transition out of four survives, and

$$\Re_0 = \left(\frac{1}{4\sqrt{3}}\frac{g_1^*}{g_1}\frac{g_e}{|g_e|}|a_1| - \frac{g_1}{|g_1|}\frac{g_e}{|g_e|}\frac{a_1}{|a_1|}\cos 2\phi - \frac{1}{4\sqrt{3}}\frac{g_1^*}{g_1}\frac{g_e}{|g_e|}|a_1|\cos 4\phi\right). \qquad (7)$$

Expressions (6) and (7) were obtained on the assumption that there is a unique direction and magnitude of the deformation. However, in real QWs the deformations can be distributed in both direction and value. To account for this, we introduce a distribution function $f(\Phi)$, which determines the probability of finding a deformation whose principal axis makes an angle $\Phi$ with the direction [110]. Then, the angle that the magnetic field makes with the [110] axis will be $\varphi=\Phi+\phi$, and all nonequivalent directions of the deformation will be included between $\Phi=0$ and $\Phi=\pi$.

Generally, the function $f(\Phi)$ may contain all harmonics. However, since the expression (5) contains only the zeroth, second and fourth harmonics, we retain just these three harmonics in the expression for $f(\Phi)$:

$$f(\Phi) = \frac{1 + C_2 \cos 2\Phi + C_4 \cos 4\Phi}{2\pi}. \qquad (8)$$

The integral $\int_0^{2\pi} f(\Phi)d\Phi = 1$, and there is a limitation imposed on the values of $C_n$, as the probability $f(\Phi)d\Phi$ must be positive at any $\Phi$.

To calculate $\Re_0$ now, it is necessary to average the radiation intensities with appropriate polarizations over the directions $\Phi$. On averaging, other harmonics which are omitted in Eq.(8) do not contribute to the polarization (this justifies the present form of the distribution function Eq.(8)). The resultant polarization $\Re_0$ can be presented in the form of Eq.(5) with $\phi$ replaced by $\varphi$, while the expressions for $\wp$ and $A_i$ take the form

$$\wp = \frac{2}{\sqrt{3}} \langle a_1 \rangle C_2$$

$$A_0 = \frac{g_1 g_1^*}{\Delta^2}\left(1 - \frac{65}{6}\langle a_1^2 \rangle + \frac{4}{3}\langle a_1 \rangle^2 C_2^2 \right) - \frac{1}{3}\frac{g_e g_1^*}{\Delta kT}\left(\langle a_1^2 \rangle - \langle a_1 \rangle^2 C_2^2\right) + 2\frac{g_1 g_1^* \langle a_1^2 \rangle}{\Delta kT}$$

$$A_2 = \langle a_1 \rangle C_2 \left(-\frac{\sqrt{3}}{2(kT)^2} g_e g_1 + \frac{(g_1^{*2} - 5g_1^2)}{3\Delta^2} + \frac{2}{\sqrt{3}}\frac{g_e g_1}{\Delta kT} + \frac{\sqrt{3}g_1^2}{\Delta kT}\right) \qquad (9)$$

$$A_4 = -\frac{g_e g_1^*}{2(kT)^2}\langle a_1 \rangle^2 C_4 - \frac{g_1 g_1^*}{\Delta^2}\left(\frac{9}{2}\langle a_1^2 \rangle C_4 - \frac{4}{3}\langle a_1 \rangle^2 C_2^2\right) +$$

$$+ 2\frac{g_1 g_1^*}{\Delta kT}\left(\langle a_1^2 \rangle C_4 + \frac{\langle a_1 \rangle^2 C_2^2}{3}\right) + \frac{\langle a_1^2 \rangle C_4 g_1 g_1^*}{\Delta kT},$$

where $(\Delta E_h, \Delta E_e) \ll kT$ and $\langle ... \rangle$ implies the averaging over a magnitude. In the converse conditions $(\Delta E_h, \Delta E_e) \gg kT$ one obtains, instead of Eq.(7),

$$\Re_0 = \left(\frac{1}{4\sqrt{3}}\frac{g_1^* g_e}{g_1|g_e|}\langle a_1 \rangle - \frac{g_1^* g_e}{|g_1||g_e|}\frac{\langle a_1 \rangle}{|a_1|}C_2 \cos 2\varphi - \frac{1}{4\sqrt{3}}\frac{g_1^* g_e}{g_1|g_e|}\langle a_1 \rangle C_4 \cos 4\varphi\right). \qquad (10)$$

The authors of Ref. [2] have shown that the fourth harmonic of polarization can appear as a consequence the cubic anisotropy of the hole $g$-factor. If that is the case, the heavy hole states split linearly in the magnetic field even without the effect of an in-plane deformation. A calculation for the weak magnetic field range $g_2 H \ll kT$, where $g_2$ is the anisotropic part of the hole g-factor in the bulk material, leads to the following expression for $\Re_0$:

$$\mathfrak{R}_0 = B_0 H^2 + B_4 H^2 \cos 4\varphi,$$

$$B_0 = \frac{G_1^2}{\Delta^2}\left(1 + \frac{G_2}{G_1}\right); \quad B_4 = -\frac{g_e G_2}{2(kT)^2} + \frac{G_1 G_2}{\Delta kT} - \frac{G_1 G_2}{\Delta^2}, \tag{11}$$

$$G_1 = g_1 + \frac{7}{4} g_2, \quad G_2 = \frac{3}{4} g_2$$

The expression of $\mathfrak{R}_0$ for the arbitrary ratio between $g_2 H$ and $kT$, and for $g_1=0$ has been obtained in Ref. [2]. In the limit $g_2 H \ll kT$, the result of Ref. [2] coincides with Eq.(11). By comparison of Eqs.(5) and (6) with Eqs.(11) one can see that the deformation results in the anisotropy of the magnetic properties in the plane of the QW, where the combination $g_1^* \eta^2 / \Delta^2$ in effect plays the role of $g_2$.

## V. Discussion

We have seen that the calculation of the linear polarization of the PL from the QW subject to a magnetic field in the plane of the QW turns out to be a sophisticated problem. Even in a low-field range, the expressions for the symmetry-allowed zeroth, second and fourth angular harmonics include many contributions and it is in general impossible to point out *a priori* which contributions will be most significant. Moreover, we have available two theoretical approaches which interpret in different ways the nature of the fourth angular harmonics of the linear polarization. While one version (Eq.(11)) includes the Luttinger parameter of the cubic anisotropy of the valence band $g_2$ (i.e., it follows Ref.[2]), the other version (Eq.(9)) does not require that parameter.

At the time, the experimental data obtained for QWs of different thicknesses and in different samples show a wide variety in the key characteristics, for instance, the amplitude of the second angular harmonics at zero field ("built-in polarization") and the ratios of the magnetic field-induced zeroth, second and fourth angular harmonics. In the present section we attempt to identify clues to the microscopic origin of the valence band spin anisotropy and to the mechanisms which actually determine the linear polarization of the PL in the QWs subject to the in-plane magnetic field.

### V.1. Low symmetry and the influence of the substrate

The most general observation for almost all of the QWs in all the samples (with rare exceptions as in Fig.4) is the non-equivalence of the [110] and [1$\bar{1}$0] directions. In other words, the angular scans of the polarization always contain the second harmonics. So it is interesting to know what factors can influence the amplitude of the second harmonics.

One can see from Fig.3 that the second harmonic is present in the angular polarization scans of the PL both from the QW and from the thick barrier layer. Obviously, if the barrier layer were characterized by the nominal point symmetry $T_d$ or $D_{2d}$, the presence of the second harmonics would be impossible. This means that the in-plane distortion that reduces the symmetry to $C_{2v}$ exists not only in the QW, but also in the barrier. Since the anisotropy of the QW heterointerfaces [18] or the anisotropic localization of excitons in the QW [19] does not affect the barrier layer, this observation implies instead a perturbation of another nature, which permeates the whole heterostructure.

Comparison with the similar results in Fig. 4 shows that if the heterostructure is separated from the substrate by an ASL buffer, the second harmonic almost disappears from the angular scans of both the QW and the barrier PL. This suggests that the non-equivalence of the [110] and [1$\bar{1}$0] directions is present through the heterostructure, the reason being traced to the buffer. This conclusion agrees with the results of Ref. [20] where the non-equivalence of the [110] and [1$\bar{1}$0] directions was revealed by the X-ray diffraction study of similar heterostructures. The interpretation involved the evolution of the misfit dislocations during the growth of the heterostructure.

*V.2. In-plane distortion: regular or random?*

Scanning of the light spot over the surface of the samples shows that the PL polarization and, thus, the in-plane distortions are homogeneous over the plane of the samples on the large (millimeter) scale. As for homogeneity on the smaller scale, one can judge it by comparison of the SFRS data with the value of the built-in polarization. The easiest way is to analyze the results for the QW with the most extreme anisotropy of the hole $g$-factor (see Fig. 2(a)). In this QW, the magneto-induced polarization of the PL contains only the second angular harmonic, so that one can be certain that the transverse $g$-factor of the hole is *completely* induced by the in-plane distortions and that the splitting of the heavy hole states in the transverse magnetic field is determined by the expression $\Delta E_h = 3g_1 H \eta / \Delta$. The splitting of the heavy hole states in a longitudinal magnetic field does not depend on the value of the in-plane distortion and equals $\Delta E_h^{long} = 3g_1^{long} H$, where $g_1^{long}$ is the hole $g$-factor in the Faraday geometry (we must distinguish between $g_1^{long}$, $g_1$ and $g_1^*$; see section IV). The experimental ratio of the values of the hole splitting in the transverse and the longitudinal magnetic fields equals ~0.2, while the theory yields for that ratio $\Delta E_h / \Delta E_h^{long} = g_1 \eta / g_1^{long} \Delta$. A simple calculation of the envelope functions of the light and heavy holes for the chosen QW shows that $g_1 / g_1^{long} \approx 2$, so that for the ratio $\eta / \Delta$ we obtain the estimated value 0.1. However, according to Eqs.(6) the ratio $\eta / \Delta$ is numerically equal to the value of the built-in polarization $\wp$, for which the direct measurement gives 0.02. The discrepancy is significant since the determination of the $g$-factors by the SFRS technique is sufficiently accurate; also the value of the built-in polarization Eqs.(6) is determined solely by the wave functions of the valence band. The latter gives an advantage as compared to the magneto-induced polarization of the PL, where a quantitative analysis requires that the properties of the electron are correctly accounted for, as well as the various specific properties of the semi-magnetic layers (magnetic fluctuations [10], heating by light [21], inhomogeneity of the exchange field of manganese ions, etc.).

The contradiction in the deduced values of the ratio $\eta / \Delta$ can be resolved by taking into consideration the fact that the SFRS experiment is sensitive only to the *value* but not the sign of the hole $g$-factor, so the regions with different local orientations of the in-plane distortions will give the same value of the Raman shift. Contrary to that, in case of the built-in polarization, regions with different orientations of the in-plane distortion will give different responses. It is the result of averaging Eqs. (6) over all possible orientations of the in-plane distortion that will determine the measured value of the built-in polarization (see Eqs. (9)). In other words, the built-in polarization will differ from zero when the probabilities of in-plane distortions along [110] and along [1$\bar{1}$0] are not equal.

Hence, we are led to conclude that submillimeter regions with different orientations of the in-plane distortion exist and that the proportions of the regions distorted along [110] and along [1$\bar{1}$0] are not balanced so that, "on average", the QW possesses effective $C_{2v}$ symmetry. This conclusion should hold true for the other QWs so, for comparison with experiment, one should use the results of the calculations in the form Eqs. (9), where averaging over an ensemble of the in-plane distortions has been performed. This has important implications in the problem of the microscopic origin of the fourth angular harmonic of the polarization.

*V.3. Analysis of contributions to the angular harmonics*

The whole set of experimental results shows that the zeroth and second harmonics are dominant. The zeroth harmonic is more pronounced in the structures with Mn throughout than in the QWs with the magnetic barriers. Within a given sample, it is stronger (as compared to other harmonics) for thinner quantum wells.

One can see from Eqs. (9), which yields the quadratic approximation in the magnetic field, that the zeroth harmonic ($A_0$) has the most numerous potential sources, all having comparably large effects. In the first term, the prefactor contains in the denominator the large value $\Delta^2$; however the first term in brackets (unity, the "Van Vleck term") is not parametrically small in $a_1$, while the next (small) item has, instead, a numeric coefficient of about 10. The second and the third terms are parametrically small in $a_1$ but, in the prefactors, the smaller quantity $\Delta kT$ enters instead of $\Delta^2$. Thus, it appears that none of the contributions can be eliminated.

For the second harmonic, it is quite natural that the common prefactor $\langle a_1 \rangle C_2$ appears in Eqs. (9), since its amplitude must become zero either if $\eta/\Delta$ tends to zero or if the moment $C_2$ in the angular distribution Eq. (8) becomes zero (i.e., if the distortions along [110] and along [1$\bar{1}$0] are equally probable). Here, we note that it is the first term that mainly determines the value of the amplitude $A_2$; this term originates from the thermal spin orientation of both electrons and holes and has a squared dependence on temperature in the denominator. The rest of the contributions to $A_2$ may be neglected.

Finally, for the fourth harmonic in Eqs. (9), similar arguments show that the first term again dominates. Only if the moment $C_4$ is totally absent in the angular distribution Eq. (8) will the contributions containing $C_2$ play a role.

We would like to note that the formulae of Eqs. (9) correspond to the low–field approximation, i.e., $\Delta E_e, \Delta E_h \ll kT$. However, this model also yields a simple result, Eq. (10), for the opposite limit $\Delta E_e, \Delta E_h \gg kT$, a condition which leads to only two levels of the four being populated. While there is no guarantee that this limit is strictly reached (the theory is valid at $\Delta E_h \ll \Delta$ only), Eqs. (9) and Eqs. (10) allow one to estimate the behavior of the polarization in a realistic domain $\Delta E_e, \Delta E_h \sim kT$. Indeed, for many QWs the experiment shows a deviation of the polarization from a quadratic field dependence as the field is increased (see Figs. 6 and 7).

It is interesting to compare Eq. (10) to Eq. (7), which does not account for the distribution of the orientations of the in-plane distortions. In Eq.(7) the second harmonic always dominates, even if in weaker field the zeroth harmonic prevails (for small $a_1$ ; see Eq. (6)). In Eq. (10), the ratio of the amplitudes of the zeroth and second harmonics can be arbitrary; it is determined by the ratios between $H$, $a_1$ and $C_2$. The experiment shows that (when we were able to apply a sufficiently large magnetic field) the angular polarization scans do not evolve towards second harmonics anyway. So, a model involving a spread of orientations of the in-plane distortions appears to describe experiment better than a model with a uniform in-plane distortion.

*V.4. Relationships between the harmonics and origin of the fourth harmonic*

We now analyze the typical observed relationships between the different harmonics of the angular scans of polarization. Consider first the signs of the principal terms in amplitudes Eqs. (9).

The built-in polarization $\wp$ was defined in Eq. (5) and its sign can be chosen to be positive, for the sake of comparison with the coefficients $A_i$ in Eqs. (9). The magneto-induced second harmonic $A_2$ has a negative sign in the first (main) term. The experiment shows that if the polarization is recorded at the low-energy part of the spectrum (see for details subsection V.5) and the magnetic field increases, then the total amplitude of the second harmonic increases in absolute value monotonically. Hence, in Eqs. (9) the sign of $\wp$ must coincide with that of $A_2$. In turn, this means that the product $g_1 g_e$ must be negative. Indeed, in our theory $g_1$ and $g_e$ have opposite signs. We have confirmed this by comparison of the optical selection rules Eqs.(1) with the well-known fact that in systems based on (Cd,Mn)Te in the Faraday geometry, the lowest-energy transition between the electron and the hole spin sublevels is optically-allowed [9]. Since $g_e$ is certainly positive when the Mn exchange term dominates over band structure effects (as proved in Fig. 9), then it follows that $g_1$ is negative, consistent with Eq. (4), in which $\beta N_0$ is negative [9].

Another experimental fact is that if one fits the angular scans of $\Re_0$ by a sum of zeroth, second and fourth harmonics, the signs of the amplitudes of the zeroth and fourth harmonics always coincide. For the estimate $a_1 \approx 0.1$ the zeroth harmonic in Eqs.(9) is controlled by the "Van Vleck" first term, which is positive in our theory. In the fourth harmonic, the dominant first term is negative; however, the product $g_1^* g_e$ is negative (since $g_1^*$ and $g_1$ have the same sign, from Eqs. (4)). The coefficient $C_4$ appears to be positive, which implies the prevalence of distortions along <110> over those along <100> (we note that the preference of directions <110> in the QWs and quantum dots based on cubic semiconductors has been reported by many authors [22]). As a result, the signs of the zeroth and the fourth harmonics coincide, substantiating the experimentally established rule.

We now generalize the data on the absolute values of the harmonics. In the experimental angular scans, the zeroth and the second harmonics were the main ones. They often had comparable values, but the ratio between them in different QWs could differ from complete absence of the zeroth harmonic to the complete absence of the second harmonic. The fourth harmonic was either not visible or smaller by an order of magnitude than the dominating harmonic.

We note that the zeroth and the second harmonics have independent values already in the model with a homogeneous in-plane distortion (Eqs.(6)): the ratio between them is controlled by the value $a_1$. In the final expressions (Eqs.(9)), the coefficient $C_2$ introduces an additional freedom in the ratio between the two harmonics. It may appear that the magneto-induced second harmonic $A_2 H^2$ cannot exceed the built-in polarization $\wp$ but, as was pointed out in Ref. [2] and as observed in experiment, the magnetic field-induced second harmonic can be much larger than the built-in polarization. This is possible within the framework of Eqs. (9) because $g_1 g_e h^2/(kT)^2$ need not be a small value; it was only required that $g_1 a_1 h/kT$ and $g_e h/kT$ were small.

In the version of the model represented by Eqs.(6), the explanation of the value of the fourth harmonic presented difficulties. Indeed, in our model the second and the fourth harmonics have a common cause, namely, the presence of the in-plane distortion of the QW and, therefore, nonzero $a_1$. Comparing the principal contributions to $A_4$ and $A_2$, one can see that the fourth harmonic must always be smaller in amplitude than the second, since the former is quadratic while the latter is linear in the small parameter $a_1$. However, we have shown one case where the fourth harmonic is not small compared to the second (Fig. 4a). No difficulty arises in the present theory of Eqs. (9), as the amplitudes of the second and the fourth harmonics are controlled by different moments of the distribution function Eq. (8) ($C_2$ and $C_4$, respectively).

To summarize, the experimental results on the fourth harmonic of the PL polarization can be explained without taking into account any cubic corrections $g_2 \sum_i J_i^3 H_i$ to the spin Hamiltonian of the valence band. Certainly, such corrections would have contributed to the amplitudes of harmonics (principally to that of the fourth harmonic [2], see also Eqs. (11)). However, it is unclear whether such a contribution will be significant, as the values of the parameter $g_2$ in CdTe and (Cd,Mn)Te are poorly known [23]. According to Ref.[24], in GaAs this parameter equals 0.04, two orders of magnitude smaller than $g_1$. For this reason, it is important to seek an alternative explanation (in which $g_2$=0) for the presence of the fourth harmonic. One should note that the concept of a distribution of in-plane distortions (Eq. (8)), which allowed us to obtain an arbitrarily large ratio of the fourth harmonic to the others, was dictated by independent reasons (see subsection V.1).

*5. Spectral dependences of the polarization*

We now consider the spectral dependences of the linear polarization of the PL and shall see that a detailed understanding of the mechanisms underlying the harmonics of the polarization angular scans will be essential; the results will underline the relevance of the preceding model.

A strong spectral dependence of the polarization is typical for those QWs where the angular scans are dominated by the field-induced second harmonic (Fig. 2(a)). Fig.1(a) shows an example of such a case, where one sign of polarization (conventionally positive) dominates on the excitonic line, but within the trion line the polarization changes sign, with the high-energy wing being negatively and the low-energy wing positively polarized.

The theoretical treatment requires further comment. Equation 9 includes all sets of the electron-to-hole recombination transitions $\Gamma_6$-$\Gamma_8$; it was implicitly assumed that the linewidths corresponding to each of the transitions are larger than any of the spin splittings. Equation 10 describes the case when only one transition (lowest in energy) contributes to the radiation. For a comprehensive description of the polarization at all wavelengths and temperatures, one should sum the four lines corresponding to four transitions $\Gamma_6$-$\Gamma_8$ in such a way that the line positions are determined by the respective combinations of the electron and the hole spin splittings, the polarization is given by formulae similar to Eq.(10), and each line is summed with a weight proportional to the populations of the respective sublevels. Such a calculation is however unfeasible in view of the unknown paramters involved and is therefore not useful in the quantitative interpretation of the experimental data.

In a literal interpretation of Eqs.(9), all the terms that depend on the spin polarization of electrons, including the principal terms of the second and fourth harmonics, must become zero for the trion state where the total spin polarization of the two electrons equals zero. However, since the *final states* of recombination of the hole with this or that electron have different energies in the magnetic field, oppositely polarized emission is produced on the two wings of the PL line. Assuming that holes populate the lower spin sublevels in each of the trions, the X⁻ emission will be constructed from two lines of equal intensities. Recombination leaving behind the electron on the upper Zeeman sublevel will result in a smaller photon energy and positive polarization, while that leaving behind the electron on the lower Zeeman sublevel will produce a larger photon energy and negative polarization (just as in Fig. 1(a)). This reasoning leads one to the important conclusion: if the PL polarization is measured for the X⁻ line, the term $g_e H/kT$ in Eqs.(9) which represents the "polarization of electrons" should be replaced by the term $g_e H/\Gamma$ (where $\Gamma$ is the PL linewidth). If the holes can populate both the lower and the upper spin sublevels, as it is observed e.g. in non-magnetic CdSe/ZnSe quantum dots [3], the spectrum of X⁻ in a magnetic field will consist of four lines, which can lead to an even more complicated spectral dependence of the polarization. The same considerations affect the exciton X but, in contrast to X⁻, the populations of the two electron spin sublevels will not be equal.

One can see that the polarization spectrum in Fig. 1(b) has the same general pattern as the spectrum in Fig. 1(a). The different feature is the presence of a spectrally-independent positive contribution that moves up the spectrum in Fig. 1(b) as a whole. This contribution is due to the zeroth harmonic of the polarization (the presence of the zeroth harmonic is proved by the angular scan for the given QW). Actually, the "van Vleck" principal term of the zeroth harmonic arises completely from the wave functions of the valence band and has no relation to the splitting of levels of electrons or holes. Therefore, it has no spectral dependence. Accordingly, in those QWs where the zeroth harmonic dominates, the polarization has almost no spectral dependence (Fig. 1(c)). We note that the absence of spectral dependence makes the magneto-induced zeroth harmonic similar to the built-in polarization. This is because they have a common origin; the built-in polarization is also due to the structure of the valence-band wave functions.

In summary, the magneto-induced second (and, most probably, fourth) angular harmonics have a sharp spectral dependence as their amplitudes change sign from one optical transition to another. Contrary to that, the zero-field second harmonic (the built-in polarization) and the magneto-induced zeroth harmonic have practically no dependence on detection energy.

### V.6. *Magnetic field dependences of the polarization*

Let us now consider one more property of the angular harmonics: the dependence of their amplitudes on the value of the magnetic field. In particular, we are concerned with the characteristic field values where the quadratic dependence of a given harmonic in the field disappears. Fig. 6(a) shows that this does not happen simultaneously for all harmonics: the zeroth harmonic remains a quadratic function of the field longer than the second harmonic.

The reason can be readily understood on the basis of Eqs.(9). The principal term of the zeroth harmonics, the "Van Vleck" term $g_1^* g_1 H^2 / \Delta^2$ originates fully from mixing of the heavy and light subbands by the magnetic field, thus it keeps its form as long as $g_1 H$ and $g_1^* H$ remain small compared to $\Delta$. The second harmonic is dominated by the orientational term related to thermalization of the electrons and holes on the corresponding Zeeman sublevels (with the proviso discussed in the previous subsection). This is why here the quadratic behavior is lost earlier; it disappears as soon as either of the splittings (electron or the hole) becomes comparable to $kT$.

It is not so easy to understand why the fourth harmonic ceases to be quadratic in field earlier than the second one (Fig.7). In the model of the present paper we did not take into account the cubic symmetry of the host crystal. However one can show that if this symmetry were taken into account, the fourth harmonics would acquire additional contributions $\sim H^4$ (not only those containing $g_2$, see Eq.(11)). As, in our theory, the terms $\sim H^2$ in the fourth harmonic contain the factor $\eta^2/\Delta^2 \ll 1$, the contribution from $H^4$ which does not contain $\eta$ may influence the result already in rather weak fields.

Finally, the complex field dependence in Fig. 6(b) is worth discussion on the basis of Eqs.(9) and (10). Obviously, the initial growth of polarization proceeds in the regime of Eqs.(9). At higher fields, the electron and hole spin splittings presumably exceed $kT$, and the system enters the regime of Eq.(10) where the polarization depends on the field only weakly (this is the plateau of the dependence). It is essential that because of the prefactor $C_2$ the amplitude of the polarization Eq.(10) remains significantly less than unity (compare with Eq.(7)). Upon a further increase of the field, the spin splitting becomes comparable to the hole subband separation $\Delta$ and the model based on perturbation theory ceases to be valid. The increase of polarization in Fig.6(b) recommences.

## VI. Conclusions

We have analyzed in detail the contributions of different symmetries to the linear polarization of the PL of QWs, as well as the physical mechanisms underlying them. The magnetic field, angular and spectral dependences of the PL polarization along with the data on the spin-flip Raman scattering were used for construction and verification of the theoretical model. We have shown that in real QWs, the effects related to the breakdown of the in-plane symmetry are essential for the linear polarization of the PL and also determine the spin splitting of the valence band states in the in-plane magnetic field. On the contrary, the "semimagnetic" nature of the QWs under study has not resulted in any qualitatively new effects (though it was useful in experiments, shifting the magnetooptical phenomena toward the smaller magnetic fields applied). Thus the conclusions of the present paper apply in full measure to non-magnetic QWs.

In most of the QWs, the zero-field PL possesses a weak linear polarization (the "built-in polarization") whose direction is linked to the crystal axes but whose value changes only weakly over the whole PL spectrum. The *mixing* of the valence band states *by the in-plane distortions* is responsible for this polarization. For a magnetic field applied in the plane of the QW, field-induced contributions to the linear polarization appear, which differ in their dependence on the angle of rotation of the crystal around the growth axis. The first contribution (the zeroth angular harmonic) corresponds to the polarization whose direction is determined by the magnetic field but which has a weak spectral dependence. This polarization mainly originates from the "Van Vleck" type term, which depends on the *mixing* of the valence band states produced *by the magnetic field*. The second contribution (the second harmonic) corresponds to the polarization whose direction is related to the crystal axes, similar to the built-in polarization, but, contrary to that, its magnitude has a strong spectral dependence. Similar to the built-in polarization, this contribution is totally due to the in-plane distortions, but, unlike the built-in polarization, it originates in the *splitting of the electron and hole sublevels* by the magnetic field. The third contribution (the fourth harmonic) has the 90-degrees symmetry and is the most debatable in nature. We have shown that a contribution of such type can be obtained in the quadratic approximation in the magnetic field without invoking terms in the valence band spin Hamiltonian which are cubic in $J$ (as was done in Ref.[2]) and, moreover, with no explicit account of the cubic elements of symmetry of the crystal.

We also addressed the question of the origin of the in-plane distortions in the heterostructures under study and discovered a correlation between the presence of such distortions in the QWs and in the barrier layers, as well as the effect of the type of substrate and of the buffer layer. A quantitative analysis of the results, and especially the inconsistency between the values of the built-in polarization and the transverse *g*-factor of holes, has led us to the conclusion that the distortions are directed randomly on a mesoscopic scale.

In general, the theory developed withstands well the experimental tests and explains reasonably well the large amount of data on the linearly polarized luminescence in CdTe/(Cd,Mn)Te and (Cd,Mn)Te/(Cd,Mg,Mn)Te QWs.

**VII. Acknowledgements**

The work has been carried out with financial support from INTAS (03-51-5266) and Russian Foundation for Basic Research, scientific programs of the Russian Academy of Science and the Science and Education Ministry of the Russian Federation. A.K. acknowledges assistance of the Russian Science Support Foundation.